%% file: LAR_ESS_NuPhys.tex
\documentclass[12pt]{article}

\usepackage{graphicx} 
\usepackage{subcaption} 
\usepackage{braket} 
\usepackage{anysize}
\usepackage{hyperref}

\newcommand\pubnumber{NuPhys2016-Saul-Sala}
\newcommand\pubdate{April 30, 2017}

\def\IFIC{Departamento de F\'isica Te\'orica and Instituto de F\'\i sica Corpuscular \\ Centro Mixto UVEG-CSIC, Valencia, Spain}
\def\support{\footnote{Work supported by the Spanish Ministerio de Econom\'ia y Competitividad and the European Regional Development Fund under Contracts No. FIS2014-51948-C2-1-P, FIS2014-51948-C2-2-P and SEV-2014-0398.}}

\def\Title#1{\begin{center} {\Large #1 } \end{center}}
\def\Author#1{\begin{center}{ \sc #1} \end{center}}
\def\Address#1{\begin{center}{ \it #1} \end{center}}

\newcommand\pubblock{\rightline{\begin{tabular}{l} \pubnumber\\
         \pubdate  \end{tabular}}}
\newenvironment{Abstract}{\begin{quotation}  }{\end{quotation}}
\newenvironment{Presented}{\begin{quotation} \begin{center} 
             PRESENTED AT\end{center}\bigskip 
      \begin{center}\begin{large}}{\end{large}\end{center} \end{quotation}}

\input econfmacros.tex
\def\be{\begin{equation}}
\def\ee{\end{equation}}
\def\bea{\begin{eqnarray}}
\def\eea{\end{eqnarray}}
\def\bear{\begin{array}}
\def\ear{\end{array}}
\def\bfig{\begin{figure}}
\def\efig{\end{figure}}
\def\bcen{\begin{center}}
\def\ecen{\end{center}}
\def\bi{\begin{itemize}}
\def\ei{\end{itemize}}

\def\raw{\rightarrow}

\def\chic{\scriptscriptstyle}

\begin{document}
\begin{titlepage}
\pubblock

\vfill
\Title{Radiative decay of heavy neutrinos at MiniBooNE and MicroBooNE~\support}
\vfill
\Author{Luis Alvarez-Ruso and Eduardo Saul-Sala}
\Address{\IFIC}
\vfill
\begin{Abstract}
The MiniBooNE experiment reported results from the analysis of $\nu_e$ and $\overline{\nu}_e$ appearance searches, which showed an excess of signal-like events at low reconstructed neutrino energies with respect to the expected background. A proposed explanation for this anomaly is based on the existence of a heavy ($\sim 50$~MeV) sterile neutrino. These $\nu_h$ would be produced by $\nu_\mu$ electromagnetic and neutral current interactions. A fraction of them decays radiatively inside the detector. The emitted photon is misidentified as an electron or positron in MiniBooNE. We have investigated the $\nu_h$ production by coherent and incoherent electroweak interactions at the MiniBooNE and MicroBooNE targets, CH$_2$ and Ar, respectively. Studying the $\nu_h$ propagation inside the detector, we obtain the energy and angular distributions of emitted photons for a choice of model parameters.  The distinctive shape and total number of photon events from this mechanism at MicroBooNE makes its experimental investigation possible. 
\end{Abstract}
\vfill
\begin{Presented}
NuPhys2016, Prospects in Neutrino Physics

Barbican Centre, London, UK,  December 12--14, 2016
\end{Presented}
\vfill
\end{titlepage}
\def\thefootnote{\fnsymbol{footnote}}
\setcounter{footnote}{0}

The paradigm of three mixing flavors of neutrinos emerges from oscillation experiments with solar, atmospheric, reactor and accelerator neutrinos in which the square-mass differences and mixing angles have been determined with ever growing precision. Nevertheless, a number of anomalies that challenge this picture has been observed. One of them, reported by MiniBooNE, has found an excess of electron-like events over the predicted background in both $\nu$ and $\bar \nu$ modes~\cite{AguilarArevalo:2008rc, Aguilar-Arevalo:2013pmq}. The excess is concentrated at $200 < E_\nu^{\mathrm{QE}} < 475$~MeV, where $E_\nu^{\mathrm{QE}}$ is the neutrino energy reconstructed assuming a charged-current quasielastic (CCQE) nature of the events.

Existing analyses struggle to accommodate this result together with world oscillation data, even in presence of one or more families of sterile neutrinos~\cite{Giunti:2013aea}, pointing at an explanation that does not invoke oscillations. It was suggested that an underestimated background from photons emitted in neutral current (NC) interactions could account for the excess~\cite{Hill:2010zy}. Indeed, the MiniBooNE detector does not distinguish between electrons and single photons. However, studies considering nuclear effects and  acceptance corrections~\cite{Zhang:2012xn,Wang:2014nat}, obtain a number of photon-induced electron-like events which is consistent with the MiniBooNE estimate.

Gninenko proposed that additional photons could originate in the weak production of a heavy ($m_h \approx 50$~MeV) sterile neutrino slightly mixed with muon neutrinos, followed by its radiative decay~\cite{Gninenko:2009ks} In Ref.~\cite{Masip:2012ke} it was pointed out that the $\nu_h$ could also be electromagnetically produced, alleviating tensions in the original proposal with other data such as those from radiative muon capture measured at TRIUMF.     

We have revisited the scenario presented in Ref.~\cite{Masip:2012ke}. We compute  coherent and incoherent $\nu_h$ production using present understanding of electromagnetic (EM) and weak interactions on nucleons and nuclei. For a more detailed analysis, we compare to the MiniBooNE excess of events in the originally measured electron energy and angle~\cite{webpage} (being the photon ones in our case) rather than in $E_\nu^{\mathrm{QE}}$. We also take into account the experimental efficiency correction available from Ref.~\cite{webpage}.

Further insight on the nature of the MiniBooNE anomaly should be brought by the currently running MicroBooNE experiment, capable of distinguishing between electrons and photons. We have also computed the number of photon events from $\nu_h$ for the target (Argon) and geometry of the MicroBooNE detector.

We have studied $\nu_h$ EM and weak production in the following processes 
\bea
\label{nuN}
\nu_\mu \,, \bar{\nu}_\mu(k) \,+ \,  N(p) &\rightarrow&  \nu_h \,, \bar{\nu}_h(k') \,+\,  N(p') \,,  \\ [.1cm]
\label{nuA}
\nu_\mu \,, \bar{\nu}_\mu(k) \,+\, A(p) &\rightarrow&  \nu_h, \bar{\nu}_h(k')  \,+\, A(p')\,,  \\ [0.1cm]
\label{nuX}
\nu_\mu \,, \bar{\nu}_\mu(k) \,+\, A(p) &\rightarrow&  \nu_h, \bar{\nu}_h(k')  \,+\, X(p')  \,.
\eea
Reaction (\ref{nuA}) is coherent while (\ref{nuX}) is incoherent; excited states $X$ include any number of knocked out nucleons  but no meson production. The considered targets are $N=p$ and $A=^{12}$C for MiniBooNE (CH$_2$),and $A=^{40}$Ar for MicroBooNE.

In the EM case, following Ref.~\cite{Masip:2012ke}, we have adopted the effective interaction
\be
\mathcal{L}_{eff} = \frac{1}{2}\mu_{tr}^i\left[\overline{\nu}_h \sigma_{\mu\nu}\left(1-\gamma_5\right)\nu_i+ \overline{\nu}_i \sigma_{\mu\nu}\left(1+\gamma_5\right)\nu_h\right]\partial^\mu A^\nu\,,
\ee
in terms of a real transition coupling $\mu_{tr}^i$; $\nu_h$ is assumed to be a Dirac fermion of mass $m_h$.  
For all the reactions under consideration, the EM amplitude can be cast as
\be
\label{ampEM}
\mathcal{M}_{\chic EM}=\frac{e\,\mu_{tr}^\mu}{2 \left(q^2+i\epsilon\right)}\,\overline{u}(k')\,q_\alpha\,\sigma^{\alpha\mu}(1-\gamma_5) u(k) \braket{Y(p')|J^{\chic EM}_\mu|N(p)} \,,
\ee
where $q=k - k'= p'-p$. EM current $\braket{Y(p')|J^{\chic EM}_\mu|N(p)}$, with $Y=p\,,A\,,X$, is the same probed in the corresponding electron-nucleus elastic scattering processes. For the nucleon, it is given in terms of electric and magnetic form factors (FF), for which we have adopted standard dipole parametrizations. For coherent scattering (\ref{nuA}), the current is proportional to the nuclear FF, obtained as the Fourier transform of the empirical charge density distribution. Finally, for the incoherent reaction we take into account particle-hole excitations in infinite nuclear matter, adapted to finite nuclei using the local density approximation.

In the weak case, the neutrino vertex has the same structure as in the Standard model, so that the amplitude
\be
\label{ampW}
\mathcal{M}_{\chic W}= - U_{\mu h} \frac{G_F}{\sqrt{2}}\, \overline{u}(k')\gamma^\mu (1-\gamma_5) u(k) \braket{Y(p')|J^{\chic W}_\mu|N(p)} 
\ee
reduces to the one for neutrino nucleus NC scattering in the limit of mixing $U_{\mu h} \raw 1$ and $m_h \raw 0$. With the weak hadronic current $\braket{Y(p')|J^{\chic W}_\mu|N(p)}$ we have proceeded as with the EM current. As usual, vector FF are related to the EM ones; for the axial FF we have adopted the conventional dipole parametrization with $M_A =1$~GeV.    

\begin{figure}[h!]
	\centering
	\begin{subfigure}[b]{0.38\textwidth}
   				\includegraphics[width=\textwidth]{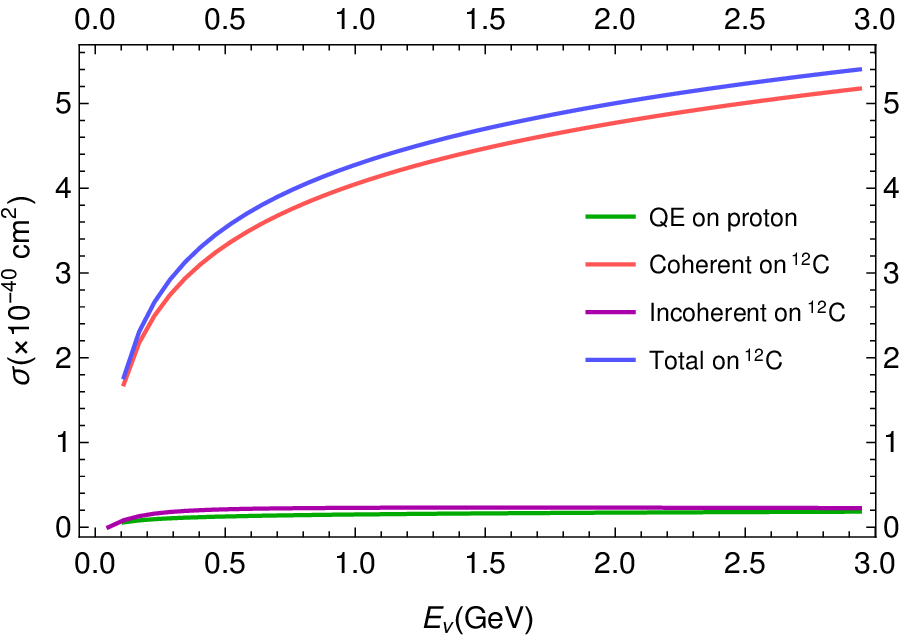}     
	\end{subfigure} $\qquad$
	\begin{subfigure}[b]{0.41\textwidth}
                \includegraphics[width=\textwidth]{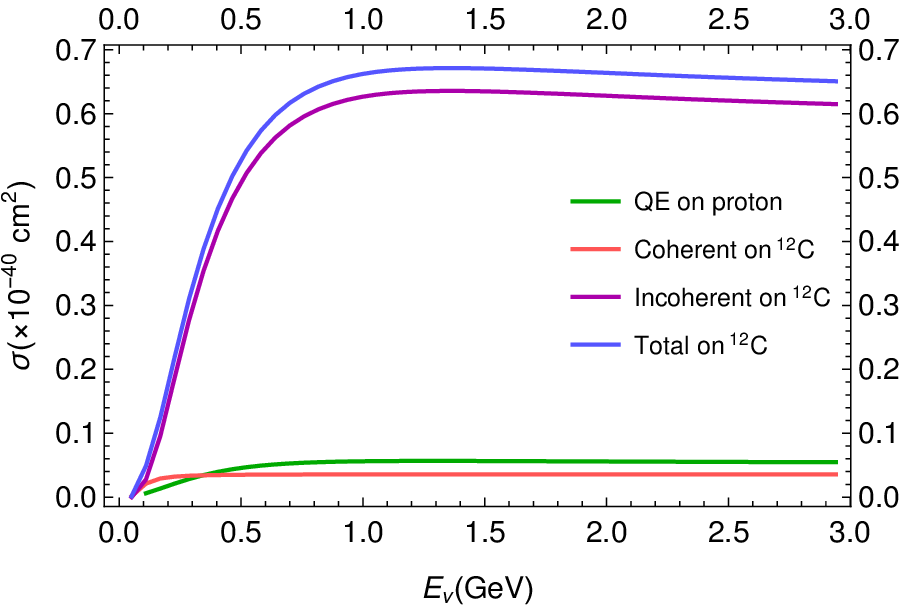}
	\end{subfigure}
\caption{Integrated cross sections for $\nu_h$ production in $\nu_\mu$-nucleus scattering by EM (left) and weak (right) interactions as a function of the incident neutrino energy.}
\label{fig:c.s.}
	\end{figure}

Our results for the integrated cross sections (cs)  on protons and $^{12}$C, obtained with $m_h = 50$~MeV, $\mu_{tr}^\mu=2.4\times 10^{-9} \mu_B.$ and $|U_{\mu h}|^2 =0.003$~\cite{Masip:2012ke},  are given in Fig~\ref{fig:c.s.}. The EM cs on $^{12}$C is dominated by the coherent mechanism while the incoherent one is suppressed by Pauli blocking at low $q^2$, where the amplitude is enhanced by the photon propagator [Eq.~(\ref{ampEM})].     
On the contrary, the incoherent reaction is the largest contribution to the weak cs. Similar features are observed for $^{40}$Ar target and also for antineutrino beams. 

We have then investigated the $\nu_h$ propagation, followed by their radiative decay inside the detector, and obtained the photon energy and angular distributions. We have taken advantage of the fact that, as pointed out in Ref.~\cite{Masip:2012ke}, the beam energies are large compared to $m_h$ and only an insignificant amount of the electromagnetically (weakly) produced heavy neutrinos have the spin against (aligned with) its momentum. Radiative decay photons are emitted predominantly in the direction opposite to the $\nu_h$ spin. The $\nu_h$ lifetime in its rest frame $\tau = 5 \times 10^{-9}$ seconds~\cite{Masip:2012ke}.

\begin{figure}[b!]
	\centering
	\begin{subfigure}[b]{0.4\textwidth}
   				\includegraphics[width=\textwidth]{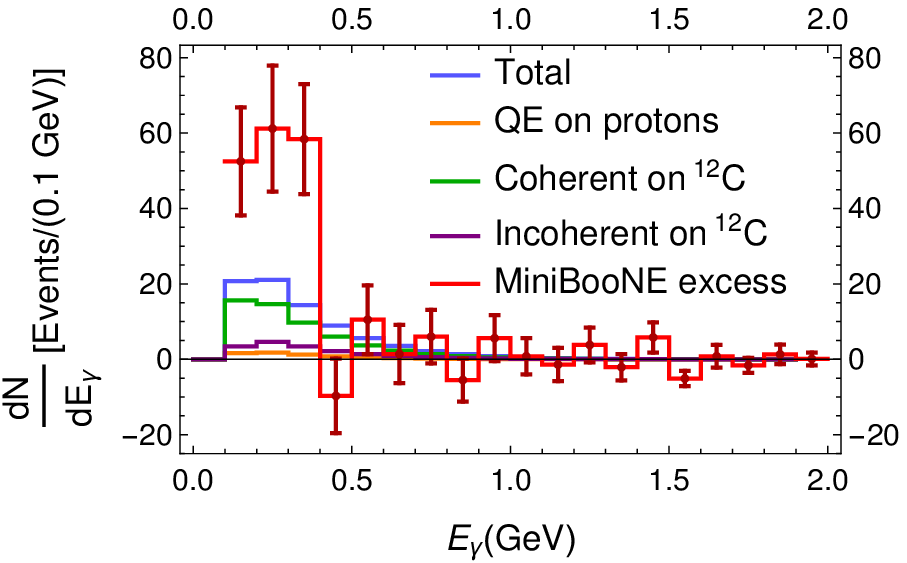}     
	\end{subfigure} $\qquad$
	\begin{subfigure}[b]{0.38\textwidth}
                \includegraphics[width=\textwidth]{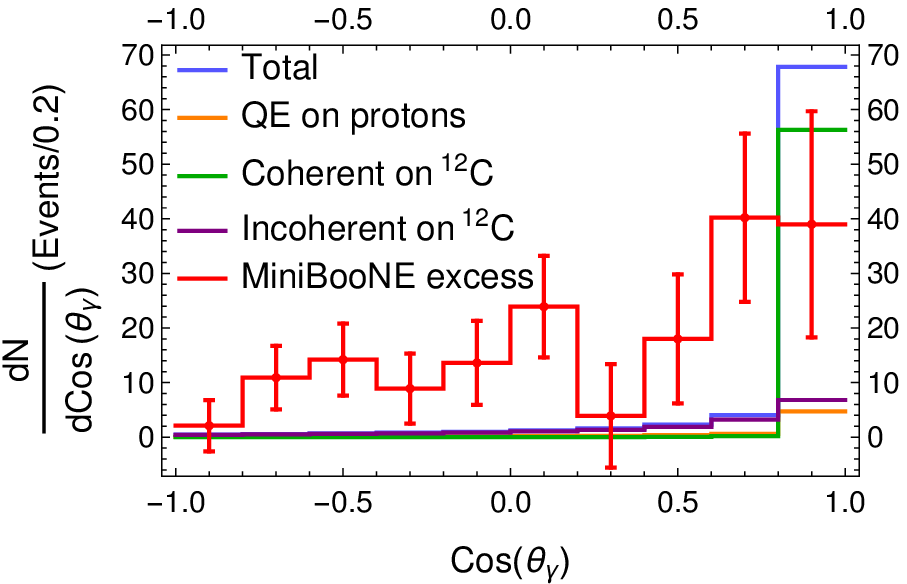}
	\end{subfigure}
	
	\begin{subfigure}[b]{0.4\textwidth}
   				\includegraphics[width=\textwidth]{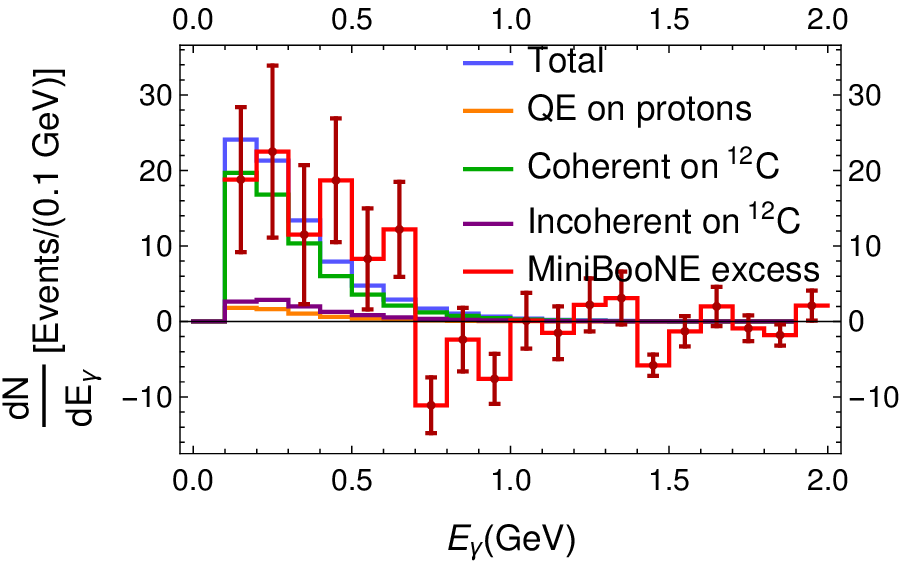}     
	\end{subfigure} $\qquad$
	\begin{subfigure}[b]{0.38\textwidth}
                \includegraphics[width=\textwidth]{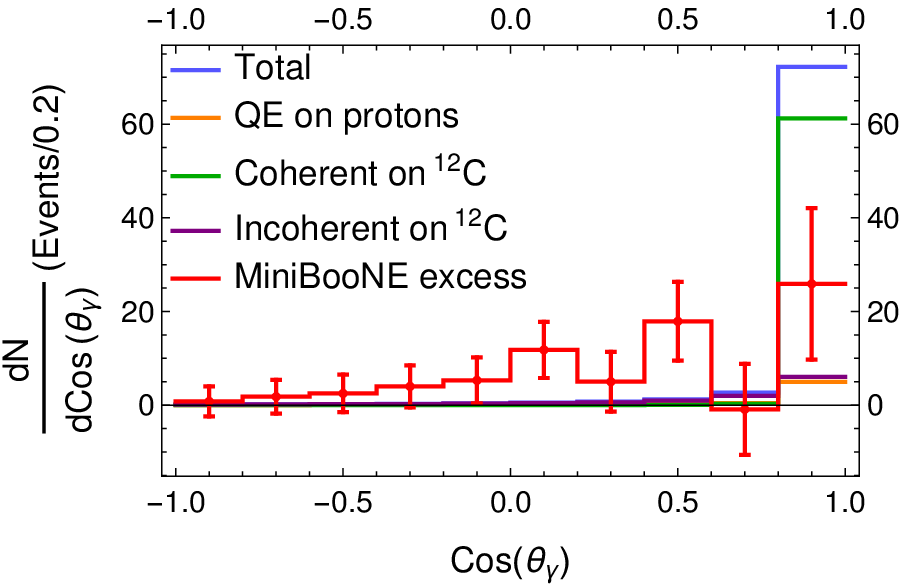}
	\end{subfigure}
\caption{Photon events from radiative decay of $\nu_h$, $\bar\nu_h$ at the MiniBooNE detector in neutrino mode (top) and antineutrino mode (bottom). Theoretical results obtained with the $\nu_h$ properties of Ref.~\cite{Masip:2012ke} are compared to the MiniBooNE excess~\cite{webpage}.} 
\label{fig:events1}
	\end{figure}

The resulting event distributions at the MiniBooNE detector for $N_{\chic POT} = 6.46\times 10^{20}$ ($N_{\chic POT} = 11.27\times 10^{20}$) in neutrino (antineutrino) modes are shown in Fig.~\ref{fig:events1}. Fluxes have been taken from Ref.~\cite{AguilarArevalo:2008yp}. 
To compare to the measured excess of events, the detection efficiency~\cite{webpage} has to be taken into account. Being energy dependent and low (at most 14~\%), its impact on the number of events is significant.
The contribution from the two protons in the target, coherent and incoherent scattering on $^{12}$C are separately shown. The number of low energy events is underestimated in $\nu$-mode, while the agreement is good in $\bar\nu$-mode. The predominantly EM coherent contribution is strongly forward peaked. This leads to a very narrow angular distribution not observed in the experiment. This result is in line with the findings of Ref.~\cite{Radionov:2013mca}.

\begin{figure}[h!]
	\centering
	\begin{subfigure}[b]{0.4\textwidth}
   				\includegraphics[width=\textwidth]{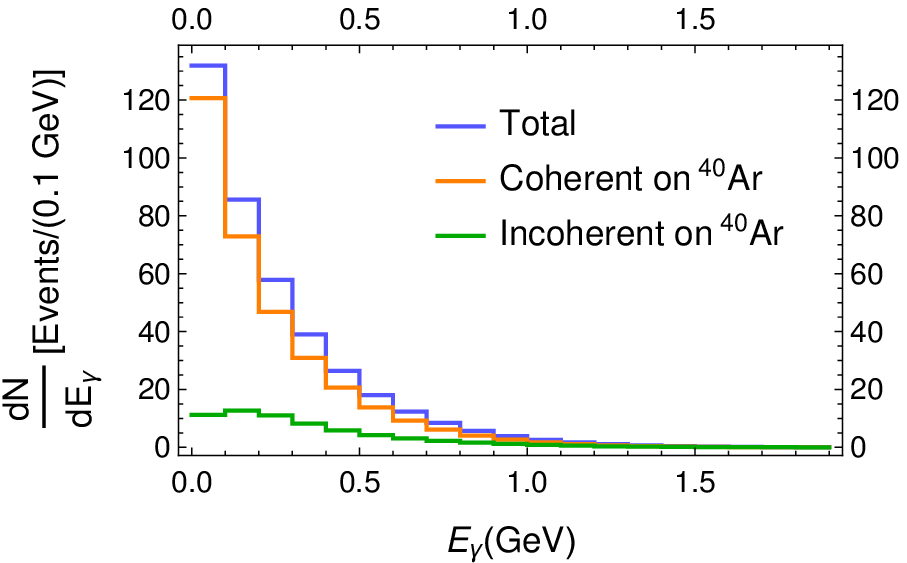}     
	\end{subfigure} $\qquad$
	\begin{subfigure}[b]{0.4\textwidth}
                \includegraphics[width=\textwidth]{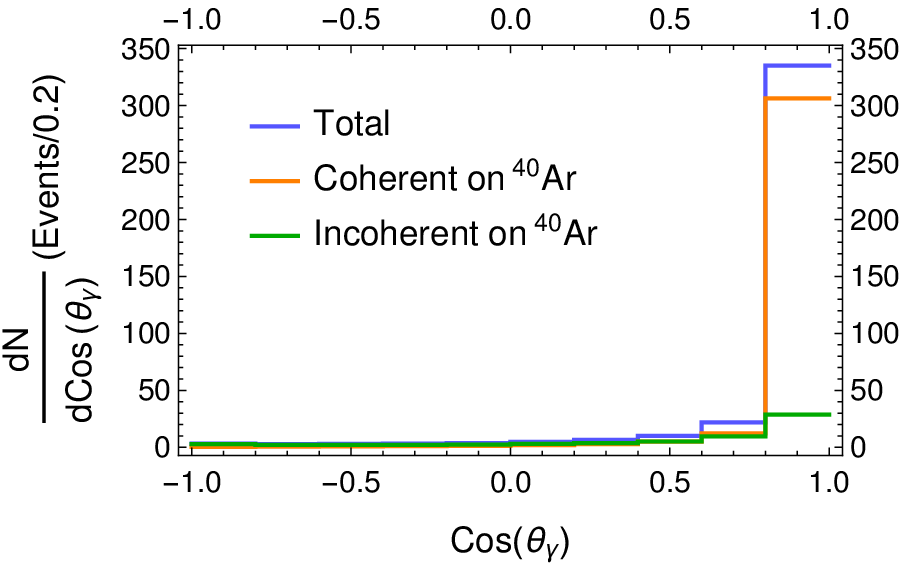}
	\end{subfigure}
\caption{Photon events from radiative decay of $\nu_h$ at MicroBooNE in neutrino mode predicted with the $\nu_h$ properties of Ref.~\cite{Masip:2012ke}.} 
\label{fig:events2}
	\end{figure}

The agreement can be improved by fitting the parameters in the allowed range. These results will be reported elsewhere~\cite{workinprog}. The radiative decay hypothesis can be tested at MicroBooNE. Our predictions for the photon distributions at this detector with the flux in $\nu$-mode~\cite{privcom} and $N_{\chic POT} = 6.6\times 10^{20}$ are displayed in Fig.~\ref{fig:events2}. The shape and number of events appears distinctive from those of conventional mechanisms.

\end{document}

%% file: econfmacros.tex
\def\beq{\begin{equation}}
\def\eeq#1{\label{#1}\end{equation}}
\def\eeqn{\end{equation}}

\def\beqa{\begin{eqnarray}}
\def\eeqa#1{\label{#1}\end{eqnarray}}
\def\eeqan{\end{eqnarray}}

\let\bar=\overbar

\def\Dslash{\not{\hbox{\kern-4pt $D$}}}
\def\dslash{\not{\hbox{\kern-2pt $\del$}}}

\def\ee{e^+e^-}

\def\msb{{\bar{\ssstyle M \kern -1pt S}}}